\documentclass[aps,preprint]{revtex4}
\usepackage{epsfig}

\begin{document}
\draft
\title{Response of Autonomic Nervous System to Body Positions: 
Fourier and Wavelet Analysis}
\author{
Aiguo Xu$^{1,2}$
\footnote{
Pesent address: Department of Physics, Yoshida-South Campus,
Kyoto University, Sakyo-ku, Kyoto 606-8501, Japan.\\
E-mail address: aiguoxu@yuragi.jinkan.kyoto-u.ac.jp}, 
G. Gonnella$^{1,2,3}$, 
A. Federici$^{1,4}$, \\
S. Stramaglia$^{1,2,3}$,
F. Simone$^{1,5}$, 
A. Zenzola$^{1,5}$, 
R. Santostasi$^{1,5}$
}
\address{
$^{1}$ TIRES, Center of Innovative Technologies for Signal Detection\\
and Processing,\\
via Amendola 173, 70126 Bari, Italy\\
$^{2}$ 
Dipartimento di Fisica, Universit\`a di Bari, \\
{\rm and} 
 Istituto Nazionale per la Fisica della Materia, Unit\`a di Bari,\\
via Amendola 173, 70126 Bari, Italy\\
$^{3}$ INFN, Sezione di Bari, via Amendola 173, 70126 Bari, Italy\\
$^{4}$ Dipartimento di Farmacologia e Fisiologia Umana - 
Universit\`a di Bari, \\
piazza Giulio Cesare 11, Bari, Italy\\
$^{5}$ Dipartimento di Scienze Neurologiche e 
Psichiatriche - 
Universit\`a di Bari, \\
piazza Giulio Cesare 11, Bari, Italy
}
\begin{abstract}
Two mathematical methods, the Fourier and wavelet transforms, were used to
study the short term cardiovascular control system. Time series, picked from
electrocardiogram and arterial blood pressure lasting 6 minutes, were 
analyzed in supine position (SUP), during the first (HD1), and the second
half (HD2) of $90^{\circ }$ head down tilt and during recovery (REC). The
wavelet transform was performed using the Haar function of period $T=2^j$ ($%
j=1$,$2$,$\cdots $,$6$) to obtain wavelet coefficients. Power spectra
components were analyzed within three bands, VLF (0.003-0.04), LF
(0.04-0.15) and HF (0.15-0.4) with the frequency unit cycle/interval.
Wavelet transform demonstrated a higher discrimination among all analyzed
periods than the Fourier transform. For the Fourier analysis, the LF 
of R-R
intervals and VLF of systolic blood pressure show more evident difference
for different body positions. For the wavelet analysis, the systolic blood
pressures show much more evident difference than the R-R intervals. This
study suggests a difference in the response of the vessels and the heart to
different body positions. The partial dissociation between VLF and LF
results is a physiologically relevant finding of this work.
\end{abstract}

\vskip 0.5 cm
\maketitle
\newpage

\section{Introduction}

\subsection{Physiological background}

Like all other animals, human body must monitor and maintain a constant
internal environment. At the same time, it must monitor and respond to the
external stimulus. Two integrated and coordinated organ systems, the
nervous system and the endocrine system, are responsible for these two
functions. The nervous system monitors and controls nearly every organ system
through a series of positive and negative feedback loops. The central
nervous system includes the brain and spinal cord. The peripheral nervous
system connects the central nervous system to the other parts of the body.
There are two subdivisions in the peripheral nervous system, the somatic and
the automnomic. The somatic nervous system includes all nerves controlling
the skeletal muscular system and external sensory receptors. The somatic
nervous system is involved in voluntary control. The autonomic nervous
system is that part of the peripheral nervous system which controls internal
organs which are not under conscious control. The autonomic nervous system
has two subsystems, sympathetic and parasympathetic. The sympathetic nervous
system is involved in response to stress conditions. The parasympathetic is
involved in relaxation. Each of these subsystems usually operates in the 
reverse of 
the other. Many organs are innervated by both of them, and their levels of 
activity are reciprocally balanced to 
maintain homeostasis.

The autonomic nervous system controls muscles in the heart. It directly
controls the cardiovascular system, both of its electrical and
mechanical properties. Via its excitation, inhibition and modulation,
it regulates both the heart rate and the propagation of cardiac
electrical activity. The autonomic nervous system controls
 a wide range of
variables, including arterial pressure, cardiac output, blood flow, and
vasomotor tone. It is responsible for the immediate response of the heart
and the blood vessels to internal noise courses as well as external 
perturbations. The former includes respiration,
digestion, emotion, and etc. 
The latter includes 
body posture, physical activity, temperature, and others.

In a healthy individual, the
beat-by-beat adjustment of hemodynamic parameters 
by the autonomic nervous system
is essential to an 
adequate cardiovascular functioning. Therefore, cardiovascular control, as
expressed by the time-dependence of hemodynamic variables, is a direct
reflection of autonomic activity. It may be used as a probe of autonomic
performance or maturation and a detector of possible autonomic malfunction.

Spectral analysis of variablity in heart rate and blood pressure
 provides a quantitative
noninvasive means of assessing the functioning of the short-term
cardiovascular control systems. Sympathetic and parasympathetic nervous
activity makes frequency-specific contributions to power spectra\cite
{science1981} of heart rate and blood pressure.

\subsection{Motivation and objectives}

Due to the regulation mentioned above, one can design experiments relating
to internal noises and/or external perturbations to investigate the
responsiveness of autonomic nervous system. This is the basic motivation and
objective of the research in this line. Many interesting results have been
achieved\cite{Review1,ConPhys,Review2,time-interval}. The present research was made
on commitment of {\it Italian Soccorso Alpino e Speleologico} 
(a group of
volunteer helpers for rescue in difficult environments) 
to study body
changes in helpers who often must go in head-down positions to rescue
wounded persons in caves or under ruins of crushed buildings. 
Specific objective of this research is to
study the response of autonomic nervous system to body positions by spectral
analysis of both heart rate variability and systolic blood pressure.

The structure of the following part of this paper is as follows: 
In Sect. II we describe the methods used in this research, which include how we
chosed the subjects, how recorded the physiological signals, how picked out
time-series for the further study, numerical tools and statistics. Section III
 gives the numerical results and physiological interpretations, which is the 
core of this paper. We draw a conclusion in Sect. IV.   

\section{Methods}

\subsection{Subjects and database}

We studied nine male subjects (33$\pm $5.7 years old) labelled as DE, GI, LO,
PI, PS, RA, SA, SC, and TA, respectively. For each subject we recorded the
electrocardiogram (ECG) and the blood pressure\ (BP) signals for consecutive
four times. The BP signals was obtained by photoplethysmographic recording 
of a sphygmogram at a finger of the right hand, maintained throughout all the 
experiment at the same hydrostatic level as the heart. 
The initial status is supine and the recording time is $6$
minutes (SUP). Then the body position of the subject was slowly, in 
about 20 seconds,  tilted to $90^{\circ }$ head down (HD). Just after 
the transition of body posture, a
second set of signals of $6$ minutes (HD1) was recorded. After that keeping
the head down position for about $10$ minutes, then a third set of 
signals of $6$ minutes (HD2) was recorded. Finally the body position 
of the subject was slowly 
recovered to supine and the fourth set of signals of $6$ minutes (REC) was
recorded.

The lower frequency limit 0.02 Hertz (Hz), corresponding to the higher 
period limit 50 seconds, is 
typical and adequate for human adults when performing short-term spectral
analysis. A generally accepted trace length should contains approximately
five full periods of the slowest investigated fluctuation\cite{trace}. Our
data length is about 360 seconds. The sampling rate is 1000 Hz. So both the
sampling rate and the trace length are acceptable.

\subsection{PQRST Sequence and time-series}

Readily recognizable features of the ECG wave pattern are designated by the
letters P, Q, R, S, and T. The wave itself is often referred to as PQRST or
QRS complex. Aside from the significance of various features of the QRS
complex, the timing of the sequence of the QRS complexes over tens,
hundreds, and thousands of heart beats is also significant. These
inter-complex times are readily measured by recording the occurrences of the
peaks of the large R waves. The number of R-R intervals within one minute is
the generally called heart rate.
From the ECG, the R-R intervals compose our
original time-series $\{\tau _i\}$, where $i$ is the index of the R peak or
the index of the R-R intervals. Similarly, from the blood pressure, the
systolic blood pressures (SBP) compose our original time-series $\{y_i\}$,
where $i$ is the index of the peak of SBP or the index of the SBP-SBP
intervals. Here we stress the original to distinguish them from the
resampled ones. Our numerical tools, Fourier transform and wavelet
transform, require that the time series be evenly distributed. From the
interval-based point of view time-series $\{\tau _i\}$ and $\{y_i\}$ are
evenly distributed, while from the time-based point of view, $\{\tau (t_i)\}$
and $\{y(t_i)\}$ are unevenly distributed. The two kinds of time-series 
describe the same physiological process in a little different ways. 
If one wants to use the
time-based time-series for spectrum analysis, he needs to resample them. In
this research the used resampling time interval is 0.25 second, which results
in a Nyquist critical frequency\cite{Recipe} $2$ Hz. One must care the
correspondence of the frequency unit and that of the used time-series. If
one uses the interval-based time-series, the frequency unit is
cycles/interval. If one uses time-based time-series, the frequency unit is
cycles/(time unit). If the time unit is second, then the frequency unit is
Hz. For typical interbeat interval sequences, the two kinds of frequency
units are correlated by a formula\cite{time-interval}

\begin{equation}
f_{time}=f_{interval}/E[\tau ]\text{,}  \label{time-interval}
\end{equation}
where $E[\tau ]$ represents expectation value of the R-R intervals,
$f_{time}$ and $f_{interval}$ are respectively the time-based and 
the interval-based frequency.
 The values of $E[\tau]$ for all subjects and the four body
 positions are shown in Table I. The mean value of systolic blood pressure is
also an important measure for the body conditions. The corresponding values
are shown in Table II.

{\bf Table I:} {\footnotesize Mean R-R intervals for different subjects and
different body positions. The time unit is second.} 
\[
\begin{tabular}{|l|l|l|l|l|l|l|l|l|l|}
\hline
subject index & 1 & 2 & 3 & 4 & 5 & 6 & 7 & 8 & 9 \\ \hline
SUP & 0.79 & 0.97 & 1.10 & 0.99 & 0.83 & 1.20 & 1.07 & 0.83 & 1.08 \\ \hline
HD1 & 0.75 & 0.97 & 1.11 & 1.05 & 0.91 & 1.05 & 1.01 & 0.81 & 1.07 \\ \hline
HD2 & 0.77 & 0.97 & 1.00 & 1.04 & 0.89 & 1.09 & 0.93 & 0.82 & 1.06 \\ \hline
REC & 0.86 & 1.13 & 1.15 & 1.07 & 0.91 & 1.21 & 1.08 & 0.96 & 1.10 \\ \hline
\end{tabular}
\]

{\bf Table II:} {\footnotesize Mean values of the systolic blood 
pressure for different
subjects and different body positions. The unit is mmHg. } 
\[
\begin{tabular}{|l|l|l|l|l|l|l|l|l|l|}
\hline
subject index & 1 & 2 & 3 & 4 & 5 & 6 & 7 & 8 & 9 \\ \hline
SUP & 122 & 110 & 106 & 122 & 118 & 108 & 120 & 121 & 122 \\ \hline
HD1 & 118 & 123 & 110 & 123 & 116 & 134 & 107 & 116 & 115 \\ \hline
HD2 & 119 & 123 & 107 & 121 & 120 & 134 & 105 & 122 & 113 \\ \hline
REC & 128 & 116 & 107 & 126 & 118 & 103 & 116 & 116 & 132 \\ \hline
\end{tabular}
\]

\subsection{Fourier transform}

The spectral analysis based on Fourier transform is by far the most used
quantitative method for the analysis of physiological signals. Moreover, it
has reached importance as a diagnostic tool. Fourier transform allows the
separation and study of different rhythms which are intrinsic for the
physiological signals. It makes simple a task that is difficult to perform
visually when several rhythms occur simultaneously.. The following is the
definition of the discrete Fourier transform used in the present research,

\begin{equation}
\tilde{x}_k=\frac 1{\sqrt{N}}\sum_{j=0}^{N-1}x_j\exp (i\frac{2\pi }Njk)\text{%
,}  \label{Fourier}
\end{equation}
where $\{x_j\}$ is the used time-series and $\{\tilde{x}_k\}$ is the
corresponding Fourier transform, $N$ is the number or length of the
time-series. With this definition, what the power spectrum shows is mean
contribution of each data in the time-series. From the statistical point of
view, the power spectrum is independent of the length of the time-series if
it is stable during the measuring time. In fact the nonstationarity 
generally exists if the measuring time is long. But the present research 
is focused on short term cardiovascular control system.
A drawback of the traditional
Fourier transform is that one can not observe the time behavior of the
specific frequency components. This defect is partially resolved by using
the 
Gabor transform\cite{Gabor}, also called short time Fourier transform, which
first modulates the time-series with an appropriate window function, then
performs the Fourier transform. The window function peaks around a given 
time and falling off rapidly, thus emphasizing the signal localized at
the central and suppressing the distant times.
If one shifts the window along the time, the
time behavior will be observable.

\subsection{Wavelet transform}

The Gabor transform still has a defect, the window width is fixed. In recent
years physiological time series have been considered in a more general
framework of fractal functions. See, e.g. \cite{Nature}. An extensively used
method in this kind of studies is the wavelet transform\cite{wavelet}, 
with which the window width is also adjustable. 
Wavelet analysis has proved to be a useful technique for analyzing signals
at multiple scales\cite{w1,w2,w3,w4,TeichPRL,PRL812388,Sebino}. It permits
the time and frequency characteristics of a signal to be simultaneously
examined, and has the advantage of naturally removing polynomial
nonstationarities.

The continuous wavelet transform maps a signal with one independent variable 
$t$ onto a function with two independent variables, $a$ and $b$. This
procedure is redundant and not efficient for algorithm implementations. A
more practical and convenient choice is to use a dyadic discrete wavelet
transform. For the sequence $\{\tau _i\}$, it is defined as follows,

\begin{equation}
W_{m,n}=\frac 1{\sqrt{m}}\sum_{i=0}^{N-1}\tau _i\psi (\frac im-n)\text{, }
\label{Wav}
\end{equation}
where the quantity $\psi $ is the wavelet basis function, and $N$ is the
length of the time-series, $m$ is scale which is related to the scale index $%
j$ by $m=2^j$, and $n$ is the translation variable. The interpretation of
wavelet coefficients depends on the shape of the basis function. Both $j$
and $n$ are nonnegative integers. The term dyadic refers to the use of
scales which are integer powers of $2$. This is an arbitrary choice. The
wavelet transform could be calculated at arbitrary scale values, although
the diadic scale enjoys some convenience of mathematical properties\cite
{time-interval,w2}. In this research we used the Haar wavelet,

\begin{equation}
\psi _{Haar}(t)=\left\{ 
\begin{array}{ll}
-1 & \text{for }0\leq t<0.5\text{, } \\ 
1 & \text{for }0.5\leq t<1\text{, } \\ 
0 & \text{otherwise. }
\end{array}
\right.  \label{Haar}
\end{equation}
which has been used in some physiological signals analyses. [For example, see
Ref. \cite{Sebino}.]
An extensively used measure devised from the wavelet transform is the
standard deviation of the wavelet coefficients as a function of scale\cite
{time-interval}:

\begin{equation}
\sigma (m)=\left[ \text{E}\left\{ |W_{m,n}(m)-E[W_{m,n}(m)]|^2\right\}
\right] ^{1/2}\text{,}  \label{sigma}
\end{equation}
where the expectation is taken over the time-series, and is independent of $%
n $.
In Ref.\cite{Sebino} the Haar wavelet is shown to be suitable for 
scale-dependent measures of $\sigma$.

\subsection{Frequency bands}

The periodogram covers a broad range of frequencies which can be divided
into bands that are relevant to the presence of various cardiac pathologies.
The power within each band is calculated by integrating the power spectral
density over the associated frequency range. The divisions of frequency
bands in the literatures are not exactly the same. A commonly used division 
in heart rate variability (HRV) is shown in Table III\cite{time-interval}, 
where VLF means the band of very low frequencies, LF means the band of low 
frequencies, HF means the band of high frequencies. The unit for this table
is cycle/interval. So the boundaries of frequency bands have a shift with the
 mean value of the R-R intervals, thus are related to specific subject and 
specific body conditions. In this research we used this division. Under the
 context without ambiguity, VLF, LF and HF are also used to denote the 
powers of corresponding frequency bands.  
The power of VLF maybe correlated to the oscillations of
blood vessels. The power of LF may reflect both the sympathetic and
parasympathetic activities. Efferent parasympathetic activity is a major
contributor to the power of HF\cite{time-interval}.

{\bf Table III:} {\footnotesize A division of frequency bands. The frequency
regime is divided into three bands,  very low frequencies (VLF), 
low frequencies (LF), and high frequencies (HF). 
The unit for this table is cycle/interval. }

\[
\begin{tabular}{|l|l|}
\hline
VLF & 0.003-0.04 \\ \hline
LF & 0.04-0.15 \\ \hline
HF & 0.15-0.4 \\ \hline
\end{tabular}
\]

\subsection{Statistics}

To evaluate the performance of different measures, one needs to do some
statistics. Comparisons among the four successive body positions, SUP, HD1,
HD2 and REC, are first performed by an ANOVA (analysis of variance) test for
repeated measures. The ANOVA test takes into account not only changes in
mean values and standard deviation, but also changes occurring in each
subject in different experimental conditions. This test is considered
significant when $p < 0.05$, where $p$ is the probability that the
data contained in the four groups of measures are not different,
but they are randomly extracted from the same pool of data. It is commonly
accepted in biological measurements that, if such a probability is less than 
$5\%$, the different groups of measures can not be considered as randomly
extracted from the same pool, but some real difference does exist among
them. This procedure should first be employed when comparisons are made
among three or more groups of measures, due to mathematical considerations
about the interdependence of the measures in different groups and their
degree of freedom. In other words, it is commonly accepted that, if we have
three or more groups of measures we can not immediately compare them
two-by-two, but an overall assessment of the variance must be made first. If
the ANOVA test gives a $p$ value less than $0.05$, special post-ANOVA tests
can be made to compare two-by-two each group of measures with the others.
If the ANOVA test gives a $p$ value larger than $0.05$, we can not get much 
meaningful information from it. But the t-test-2 analysis still can be used 
to compare measures of each two of the four groups. 

T-test-2 is a hypothesis testing for the difference in means of
two samples. We used the software, MATLAB, for these statistic computations. 
For two samples, ${\bf x}
$ and ${\bf y}$, t-test-2 in MATLAB presents three quantities,
{\it h}, {\it significance} and {\it ci}. The value of {\it h} is 
a t-test to determine
whether two samples from a normal distribution in the samples could have the
same mean when the standard deviations are unknown but assumed equal. The
result is $ h = 1 $ if you can reject the null hypothesis at the 0.05
significance level, and is $ h = 0 $ otherwise. In this research, $h=1$ 
means that significant difference is found from different body positions. 
The {\it significance} is the $p$-value associated with the T-statistic
\begin{equation}
T=\frac{{\bf \bar{x}}-{\bf \bar{y}}}{s\sqrt{\frac 1n+\frac 1m}}\text{, }
\label{Ts}
\end{equation}
where $s$ is the pooled sample standard deviation, $n$ and $m$ are the
numbers of observations in the ${\bf x}$ and ${\bf y}$ samples, and 
${\bf \bar{x}}$ and ${\bf \bar{y}}$ are the corresponding expectation 
values. The value of 
{\it significance} is the probability that the observed value of $T$ could
be as large or larger by chance under the null hypothesis that the 
mean of ${\bf x} $ is equal to the mean of ${\bf y}$. The quantity 
{\it ci} is a 95\% confidence
interval for the true difference in means. In fact in present research, 
the t-test-2 analysis was done for any value of $p$ from the ANOVA. 
In the case of the ANOVA $ p < 0.05 $, t-test-2 plays a similar role 
to that of the post-ANOVA.

\section{Results and discussions}

\subsection{Fourier analysis}

Before the Fourier transform, we first used the Hamming window function%
\cite{hamming} to modulate the time-series, then subtracted the mean value 
and rescaled the time-series so that the mean value is zero and the 
standard deviation is $1$. The powers of VLF, LF and HF are averaged 
over different shifts of the window function. For Fourier transform of 
the resampled time-series, the used window width
is $256$ seconds ($1024$ points). For Fourier transform of the original 
interval-based time-series, the used 
window width is $256$ intervals.

Figure 1 shows the power spectra of a subject for the consecutive four
body status. Here we used the resampled time-based time series, so 
the frequency 
unit is Hz. In each case, the solid line is for the R-R intervals and the
dashed line is for the systolic blood pressure. The status of the body
position is also shown in the inset of each figure. It is clear that (i) the 
most of power is localized with the frequency regime $0<f<0.4$,
(ii)the contribution of different frequency components varies with 
different body positions.
For the R-R intervals of this subject, the power in the low frequency part
decreased during the transition from SUP to HD1, and increased from HD1 to
HD2. The behavior of the systolic blood pressure are not exactly the same as
that of the R-R intervals. The power spectra for different subjects show 
evident difference which may due to personal conditions as well as 
respiratory status. That is beyond the scope of the studies described 
in this paper.
What one cares is not the specific values, but the mean value and its 
statistics. The power spectra from the original interval-based 
time-series give similar information.

Figure 2 shows the mean values
of VLF, LF and HF for different body positions, where the first column is
for the R-R intervals and the second column is for the systolic blood
pressures. The error bars are also shown for comparison. These results are
from the resampled time-based time-series. From the two figures, it is clear
that the VLF for both the R-R intervals and the systolic blood pressure has
a significant decrease when the body position changes from supine to 
$90^{\circ}$ head down, while decrease of LF for both is less evident. A
physiological interpretation is thus: when the body position changes from
supine to head down, due to the weight of the column of blood lending from
feet to the heart and the neck, the arterial pressure sensors, which are in
the aorta near the heart and in the neck, get a strong stimulation. This is
very similar to a sudden hypertension and induces a negative feedback
response. The sympathetic activity on the heart and the vessels is
inhibited, while the parasympathetic activity on the heart is increased. The
inhibition of sympathetic activity induces different effects: 
(i) decreasing the
ability of the heart wall to increase its elastance during systole, which
tends to decrease the systolic blood pressure; (ii) a dilation of vessels
which decreases the Poiseuille resistance to the blood, which tends to
lengthen the R-R intervals. The parasympathetic activity does not
significantly affect the resistance of vessels. So for both the R-R
intervals and the systolic blood pressures, the decrease in the variability
of very low frequency components can be connected to the decrease in
physiological very slow oscillations in sympathetic activity. In HD1 status
those very slow oscillations seem to be heavily dampened by the feedback
effects elicited by the hydraulic mass which is stretching pressure
receptors. It is also very interesting that these effects is not so evident
in the variability of low frequency components of R-R intervals and systolic
blood pressures. This could suggest that systems whose oscillations have
longer time constants are more influenced by changes in hydraulic
dislocations than the more rapidly oscillating systems. The contribution of
a longer oscillation period is mainly from the peripheral vessels, whose
slow oscillations in resistance affect not only blood pressures, but also,
in a reflex way, the R-R intervals.

After the SUP-HD1 transition the changes are reduced in the HD2 status. This
agrees with a progressive decay in the arterial pressure receptors
sensitivity, which is known to be significantly reduced after 10-15 minutes
 of
application of a constant mechanical stimulus. However, during the HD2
period, a great volume of blood has been transferred from the more elevated
parts of the body (for example, legs) to the thorax, so that the recovery 
from the HD2 status to the REC has the effect to suddenly displace this
hydraulic mass towards the vasodilated vessels of the legs. This mimicks a
sudden hypotension as in blood losses due to hemorrhage. The sympathetic
activity is enhanced as if it should counteract the effects of a 
hemorrhage. So the variability in very low and low frequency components of
R-R intervals becomes higher than in HD1 status, and the variability of low
frequency components also becomes higher than in HD2 status. This again
suggests a difference in the response of the vessels (mainly VLF, due to
longer time constant of their oscillations) and the heart (mainly LF, due to
shorter time constants of heart rate and contractility oscillations).

Because the low frequency oscillations are influenced by both sympathetic
(vessels and heart) and parasympathetic (heart only) systems, it seems that
very low frequency variability is more greatly dampened and ``stunned'' by
head-down position than the low frequency variability. That is to say, the
low frequency components of systolic blood pressure recovers more promptly
during the REC period than the very low frequency components. This behavior
suggests that the recovery of systolic blood pressure is mainly supported by
recovery in heart rate and contractility, but not in peripheral vessels
resistance whose VLF power remains low.

For the systolic blood pressure, the variability of high frequency
components in both the HD and the REC status are higher than that in the
initial supine position. This can only be a parasympathetic effect (perhaps
influenced by an increase in respiratory activity in head-down position).
For the R-R intervals, the behavior is not exactly the same.

For the Fourier analysis, Table IV shows the $p$ values from the ANOVA
analysis for the three measures, VLF, LF and HF. All the $p$ values are 
larger than $0.05$. So this table suggests that
for any one of the three measures, among the four successive body positions, 
no one shows significant difference from the other three under 
the significance level $0.05$. The R-R intervals and systolic blood pressures
 are facing the same situation at this point. But the LF of R-R intervals 
and the VLF of systolic blood pressures are expected to be observed more 
difference because their $p$ values are relatively smaller. Let us go to 
the t-test-2 analysis.  
Figure 3 shows the t-test-2 statistic results for any two of the four 
different body positions. The first column is for the R-R intervals and the
second column is for the systolic blood pressures. From top to bottom, the
corresponding frequency bands are VLF, LF and HF. The position indexes 1, 2,
3, 4, 5, 6 correspond to SUP-HD1, SUP-HD2, SUP-REC, HD1-HD2, HD1-REC,
HD2-REC, respectively. We found that all the $p$ values from the t-test-2 
are larger than $0.05$, so the corresponding values of $h$ are all zero. But 
 from the LF of R-R intervals, two pairs of body-positions, HD1-REC and 
HD2-REC, show relatively smaller $p$ values. That tells us that the REC 
shows evident difference from both the HD1 and the HD2 in the LF of the 
R-R intervals, which is consistent with what we got and understood 
from Fig. 2.  From the VLF of systolic blood pressure, SUP-HD1 and HD1-HD2 
show relatively small $p$ values, which also confirms again what we got
 from Fig. 2.

{\bf Table IV: }{\footnotesize The ANOVA $p$ values 
of different measures of the power spectra. Results for the R-R intervals
and for the systolic blood pressures are shown.}
\[
\begin{tabular}{|l|l|l|l|}
\hline
& VLF & LF & HF \\ \hline
R-R & 0.62 & 0.10 & 0.76 \\ \hline
SBP & 0.18 & 0.63 & 0.53 \\ \hline
\end{tabular}
\] 

\subsection{Wavelet analysis}

In this section we mainly describe the wavelet analysis on the original
interval-based time-series. The wavelet function is the Haar wavelet. The
used the scale $m=2^j$. Since the used window width is $1$, the scale 
index $j$ approximately corresponds the frequency $1/2^j$ cycle/interval. 
From Table I we know that the mean R-R interval of the nine subjects is 
$0.98$ second in SUP, is $0.97$ second in HD1, is $0.95$ second in 
HD2 and is $1.05$
second in REC. So we can approximately evaluate the frequencies
corresponding to different scale indexes. The correspondences are shown in
Table V.
Figure 4 shows the standard deviation $\sigma$ of the wavelet coefficients
as a function of the scale index $j$ and the subject index. The scale
indexes are shown in the inset of the figure. For the same subject the value
of $\sigma $ varies with the scale, and for the same scale the value of 
$\sigma $ also oscillates with specific subject. This result is for the 
initial SUP. 

To get a statistical information on  how $\sigma $ behaves
as a function of specific scales, Fig.5 shows the mean value of $\sigma $ 
of the R-R intervals averaged over different subjects. The results for 
the four body positions are shown in order. The corresponding error-bars 
are also shown. Figure 6 shows the same quantities but for the systolic 
blood pressures. During the SUP period, the
values of $\sigma $ for both the R-R intervals and the systolic blood
pressures show a similar behavior. They increase with the used scale. 
Let us understand this result. A larger scale corresponds to an   
oscillation with lower frequency. A larger value of $\sigma$ means that 
the strength of the oscillation with a given frequency changes more 
significantly with time. This result suggests that the organs in charge 
of lower frequency oscillations are more easily affected by other 
disturbance, so the corresponding oscillations show more nonstationarity 
with time. This figure also suggests that the nonstationarities of nearly 
all the frequency components denoted by $j$ were lower in the HD1 and the 
HD2 periods than in the supine period. In other words, the corresponding 
oscillations are more stable in the HD1 and the HD2 than in the 
SUP. From the physiological side, it seems that when the body is 
receiving a constant mechanical stimulus, some other weak internal noise
 or external disturbance is less effective in affecting the short term 
cardiovascular control system.   
 Table VI shows the 
significance values from the ANOVA analysis. Figures 7 and 8 show,
respectively, the t-test-2 results for the R-R intervals and the 
systolic blood pressures. The correspondences between the position 
index and the position-pairs are the same as those in Fig.3. 
For the R-R intervals, all the ANOVA $p$ values are greater than $0.05$,
while from the t-test-2 we still can find evident difference for three sets
of body positions, SUP-HD1, HD1-HD2, HD1-REC, from scale $j=5$, and can find
evident difference from SUP-HD2 from the scale $j=6$, even though the
t-test-2 $p$ values are larger than $0.05$. These results are
 consistent with those of the Fourier analysis. 
For the systolic blood pressures, all the ANOVA $p$ values are small. That is
to say, at least one of the four is evidently different from the other
three. For the scales with $j=1,2,3,4$, $p<0.05$. The further t-test-2 gives
more detailed information. Under the condition of t-test-2 $p<0.05$, (i)
when $j=1$, evident difference is found in SUP-HD1; (ii) when $j=2$ or $4$,
evident differences are found in SUP-HD1, SUP-HD2, and HD2-REC, (iii)when $%
j=3$ or $5$, evident differences are found in SUP-HD1 and SUP-HD2, (iv)when $%
j=6$, evident differences are found in SUP-HD1 and HD2-REC.

It should be mention that we also used the original interval-based
time-series for Fourier analysis and used the resampled time-series for
wavelet transform. But these two treatments are less effective in
distinguishing different body positions.

{\bf Table V: }{\footnotesize The approximate correspondences between the
scale indexes and component frequencies. The unit of frequency here is Hz.
The correspondence changes with subject and body positions because 
the mean R-R interval changes. }
\[
\begin{tabular}{|l|l|l|l|l|l|l|}
\hline
scale index & 1 & 2 & 3 & 4 & 5 & 6 \\ \hline
SUP & 0.510 & 0.255 & 0.128 & 0.064 & 0.032 & 0.016 \\ \hline
HD1 & 0.515 & 0.258 & 0.129 & 0.064 & 0.032 & 0.016 \\ \hline
HD2 & 0.526 & 0.263 & 0.132 & 0.066 & 0.033 & 0.016 \\ \hline
REC & 0.476 & 0.238 & 0.119 & 0.060 & 0.030 & 0.015 \\ \hline
\end{tabular}
\] 

{\bf Table VI: }{\footnotesize Significance\ values from the ANOVA for
different scales of the wavelet transform.}
\[
\begin{tabular}{|l|l|l|l|l|l|l|}
\hline
scale index & 1 & 2 & 3 & 4 & 5 & 6 \\ \hline
R-R & 0.726 & 0.799 & 0.561 & 0.954 & 0.171 & 0.416 \\ \hline
SBP & 0.047 & 0.019 & 0.006 & 0.021 & 0.075 & 0.093 \\ \hline
\end{tabular}
\]

\section{Conclusion}

We used two methods, the Fourier and wavelet transforms, to analyze arterial
blood pressure and heart rate temporal series recorded in four successive
body positions. The $90^{\circ }$ head-down position is used in this research. 
For the Fourier analysis, three measures, VLF, LF and HF, were used. For the
 wavelet analysis, the standard deviation of the wavelet coefficients for 
different scales were used. 
The two mathematical methods are shown to be complementary to study given
physiological signals. Time series lasting 6 minutes were analyzed with both
methods in supine position (SUP), during the first (HD1), and the second
half (HD2)of head down tilt and after recovery (REC). The wavelet transform
was performed using the Haar function of period $T=2^j$ ($j=1$,$2$,$\cdots $,%
$6$) to obtain wavelet coefficients. Power spectra components were analyzed
within three bands, VLF (0.003-0.04), LF (0.04-0.15) and HF (0.15-0.4 ) with
the frequency unit cycle/interval. In distinguishing two different body
positions, the resampled time-based time-series work more effectively if we
use the Fourier transform, while the original interval-based time-series
work more effectively if we use the wavelet transform.

Power spectra mean values showed that (i) the HF components of blood
pressure have an increases during tilt and have a decrease during recovery;
(ii) LF components of R-R intervals and systolic blood pressures increase in
recovery with respect to the tilt period; (iii) VLF components of the R-R
intervals and systolic blood pressures were lower in HD1 than in supine
position, but increased in HD2; (iv)in recovery period VLF components of R-R
intervals return nearly to the supine value, but VLF components of blood
pressure decrease during recovery with respect to during both the supine and 
the HD2.
Wavelet transform demonstrated a higher discrimination among all analyzed
periods than the Fourier transform, especially for the systolic blood 
pressures.  The oscillations of each frequency 
component denoted by $j$ of the R-R intervals and the systolic blood
pressures were lower in HD1 and HD2 with respect to in SUP. For the Fourier
analysis, LF of R-R intervals and VLF of systolic blood pressure show more
evident difference for different body positions. For the wavelet analysis,
the systolic blood pressures show much more evident difference than the R-R
intervals. These data suggest a difference in the response of the vessels
and the heart to different body positions. So the R-R intervals and the 
systolic blood pressures are also complementary in reflecting the response 
of autonomic nervous system. The partial dissociation between 
VLF and LF results is a physiologically relevant finding of this work.

As reported above, this research was committed to assess potential risk
factors due to cardiovascular changes in helpers working in head-down
position to rescue persons entrapped in difficult
environments. Physiological oscillations of SBP and R-R intervals in the
VLF, LF and HF bands are due to the time constants of the feedback loops
regulating these parameters. On the background of a proper amount of
ongoing oscillatory activity, the feedback regulation of cardiovascular
system is more easily allowed to compensate for the effects of external
or internal abrupt perturbations (accelerations, environmental
temperature, emotions, fatigue, etc.) on arterial pressure 
\cite{science1981,Review1,ConPhys,Review2,time-interval,trace}. In
physiological terms, if cardiovascular feedback control loops were not
continuously oscillating, they would buffer perturbations in a less
ready, adaptive and effective way, as it often occurs, for instance, in
diabetes and other pathologies affecting the fibers of the autonomic
nervous system innervating the heart and the vessels.

The dampening in SBP and R-R oscillations we have found in head-down
position likely reflects limitations in adaptive properties, readiness
and effectiveness of the cardiovascular control systems, which arises
when the blood pressure sensors in aorta and carotid artery are
overcharged by the mass of the blood column unusually accelerated from
the feet towards the thorax and the neck. Another point to consider is
that such a dampening mainly occurs in the VLF band for SBP, which
largely depend on changes in capacitance and resistance of arteries, but
it mainly occurs in LF band if R-R intervals are considered, which
clearly reflects heart period. To efficiently transfer hydraulic power
from the heart to the arteries, a proper coupling is required between
the heart period and the mechanical time constant of the elastic
arteries, which depends on their capacitance and resistance. An
uncoupling between the control mechanisms of these two parameters in
head-down position is suggested by the finding that SBP and R-R
intervals oscillations are affected by changes in two different bands of
frequency. Such an uncoupling can occur, for instance, in pathologies of
the autonomic nervous system, in heart failure and in shock syndrome.

In conclusion, cardiovascular feedback control mechanisms might be more prone to perturbations due to mental stress, physical effort and fatigue when helpers are working  in head-down position, and the mechanical efficiency of the coupling between the heart and the arteries could be reduced, so that a careful selection of the helpers must be performed before allowing them to work in head-down position, to exclude persons showing signs of anomalies in their cardiovascular control system, which could be exacerbated in such a body position. This work suggests that wavelet analysis should be employed together with the more widely used FFT analysis of cardiovascular parameters, for a better discrimination of potential risk factors in the selection of helpers.


{\bf ACKNOWLEDGMENT} Authors wish to thank Italian "Soccorso Alpino e Speleologico" for the
technical and financial support to this work.

\newpage
\begin{figure}[tbp]
\centerline{\epsfig{file=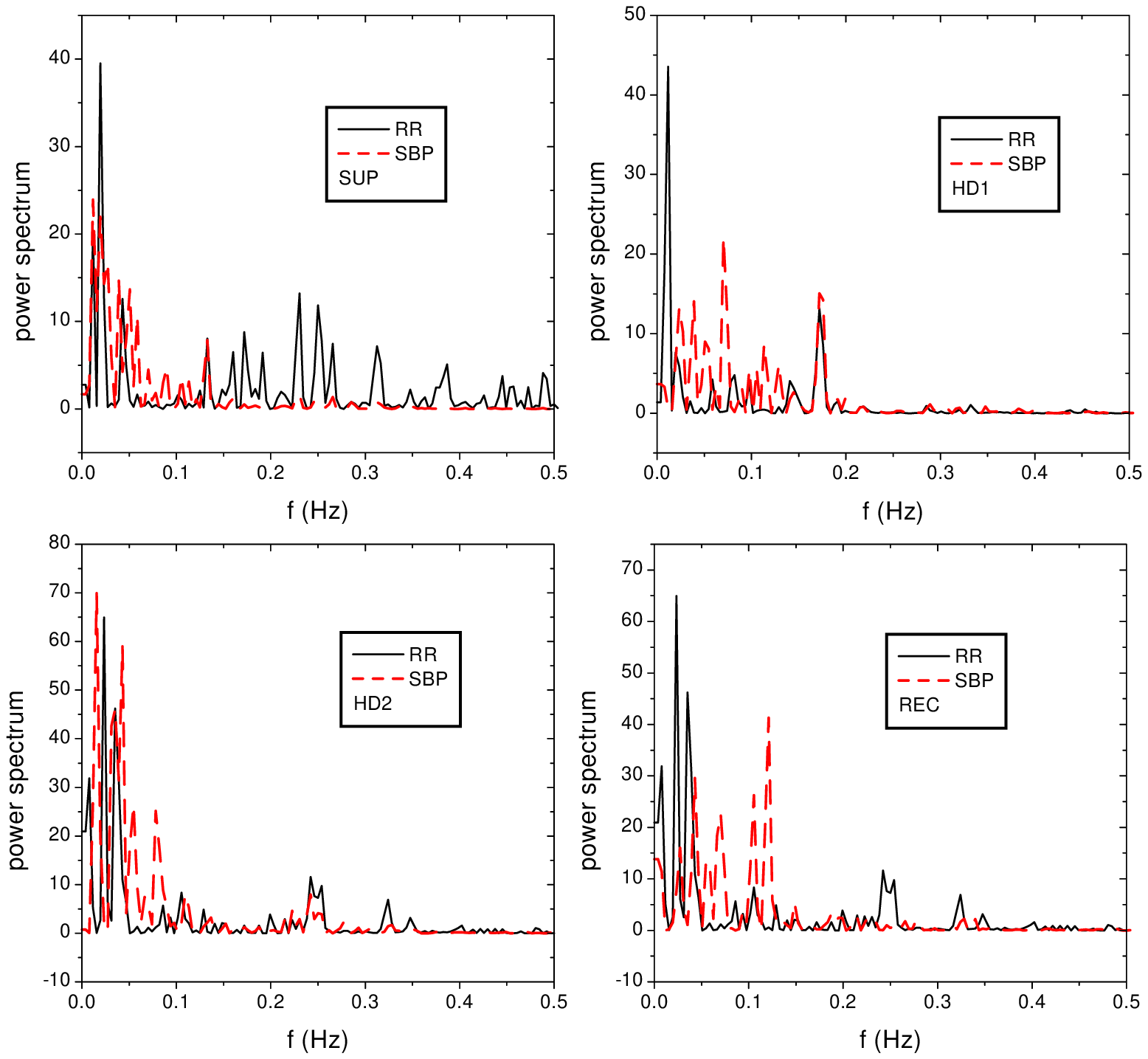,
bbllx=38 pt,bblly=321 pt,
bburx=532 pt,bbury=768 pt,
width=0.9\textwidth,clip=}}
\caption{Power spectra of the R-R intervals and the systolic blood pressures
for the subject ``DE'' for the four seccessive body positions. The 
corresponding body positions are shown in the insets of the figure.}
\label{fig_1}
\end{figure}

\newpage
\begin{figure}[tbp]
\centerline{\epsfig{file=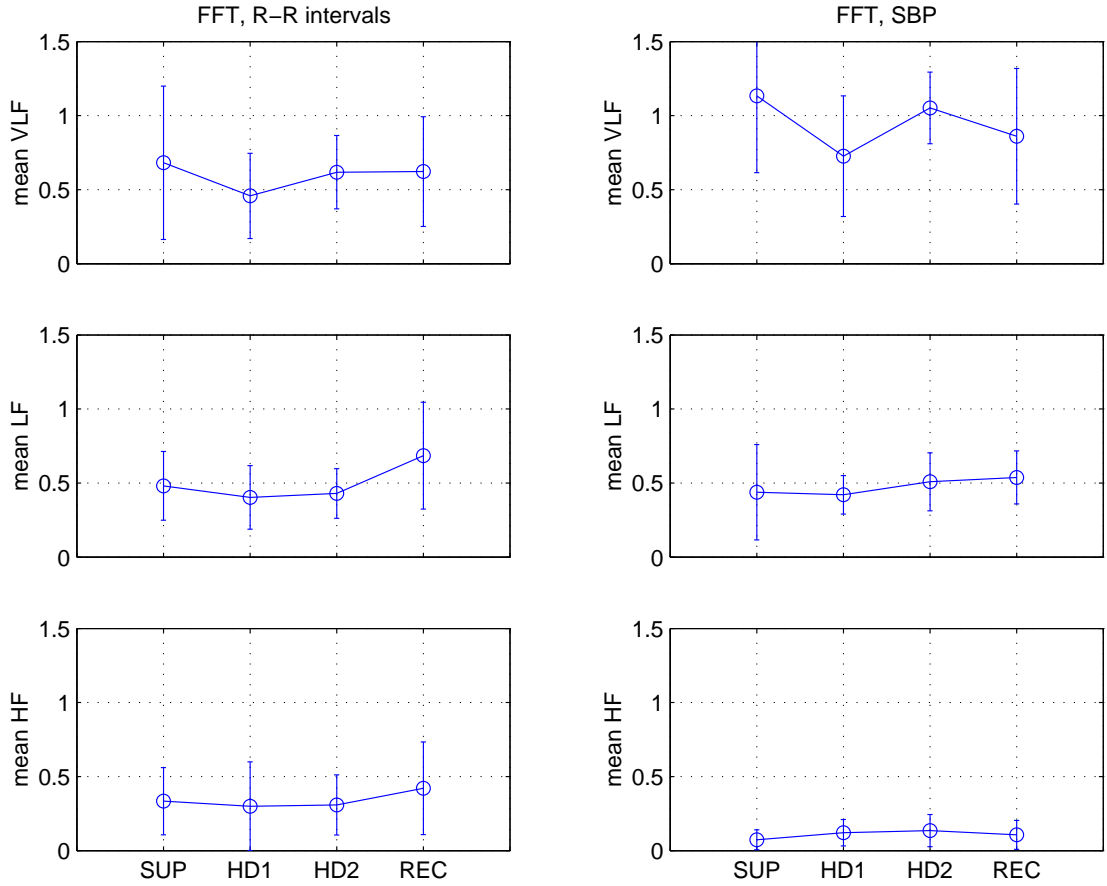,
bbllx=56 pt,bblly=210 pt,
bburx=542 pt,bbury=600 pt,
width=0.9\textwidth,clip=}}
\caption{
Mean values of VLF, LF and HF with error-bars. The first column is for the
R-R intervals and the second column is for the systolic blood pressures.
The same scales are used in each case for the convenience of comparison.
}
\label{fig_2}
\end{figure}

\newpage
\begin{figure}[tbp]
\centerline{\epsfig{file=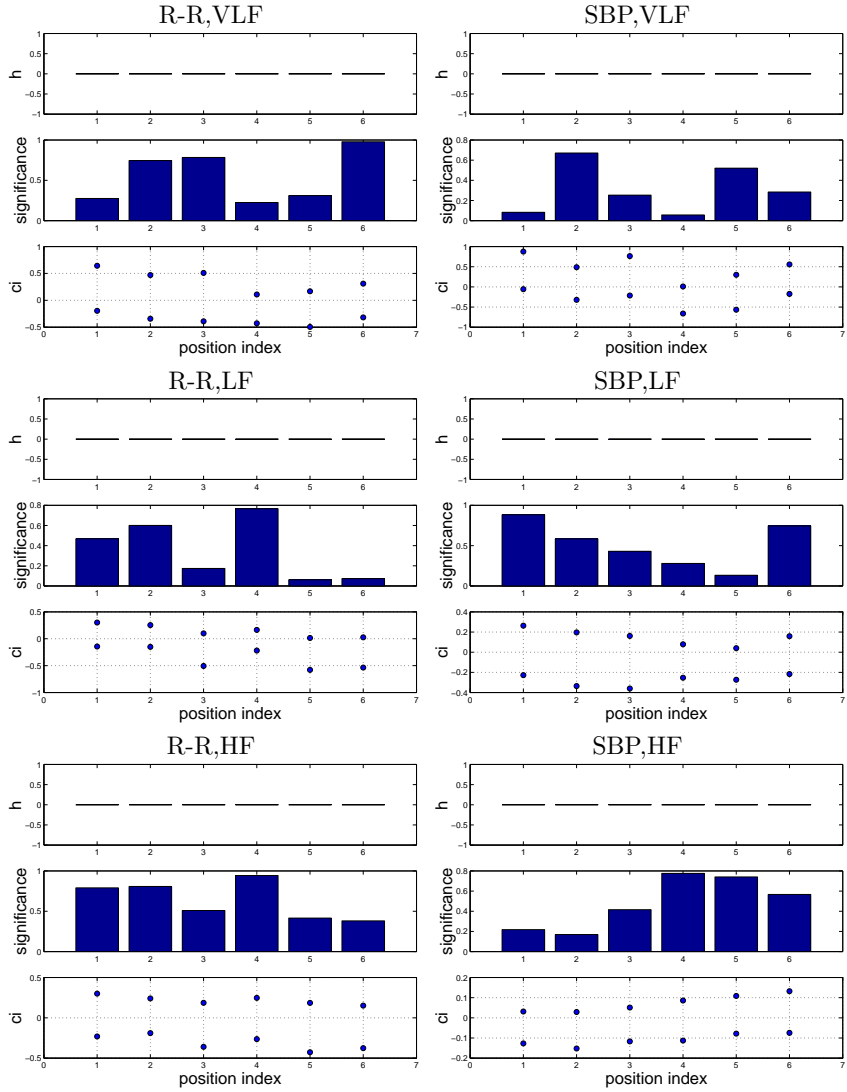,
bbllx=37 pt,bblly=211 pt,
bburx=475 pt,bbury=678 pt,
width=0.9\textwidth,clip=}}
\caption{
T-test-2 analysis results of VLF, LF, and HF for any position-pair of the
four body positions. The first column is for the R-R intervals and the
second column is for the systolic blood pressures. The position indexes
 1,2,3,4,5,6 correspond to the position-pairs SUP-HD1, SUP-HD2, SUP-REC, 
HD1-HD2, HD1-REC,
HD2-REC, respectively.  
}
\label{fig_3}
\end{figure}

\newpage
\begin{figure}[tbp]
\centerline{\epsfig{file=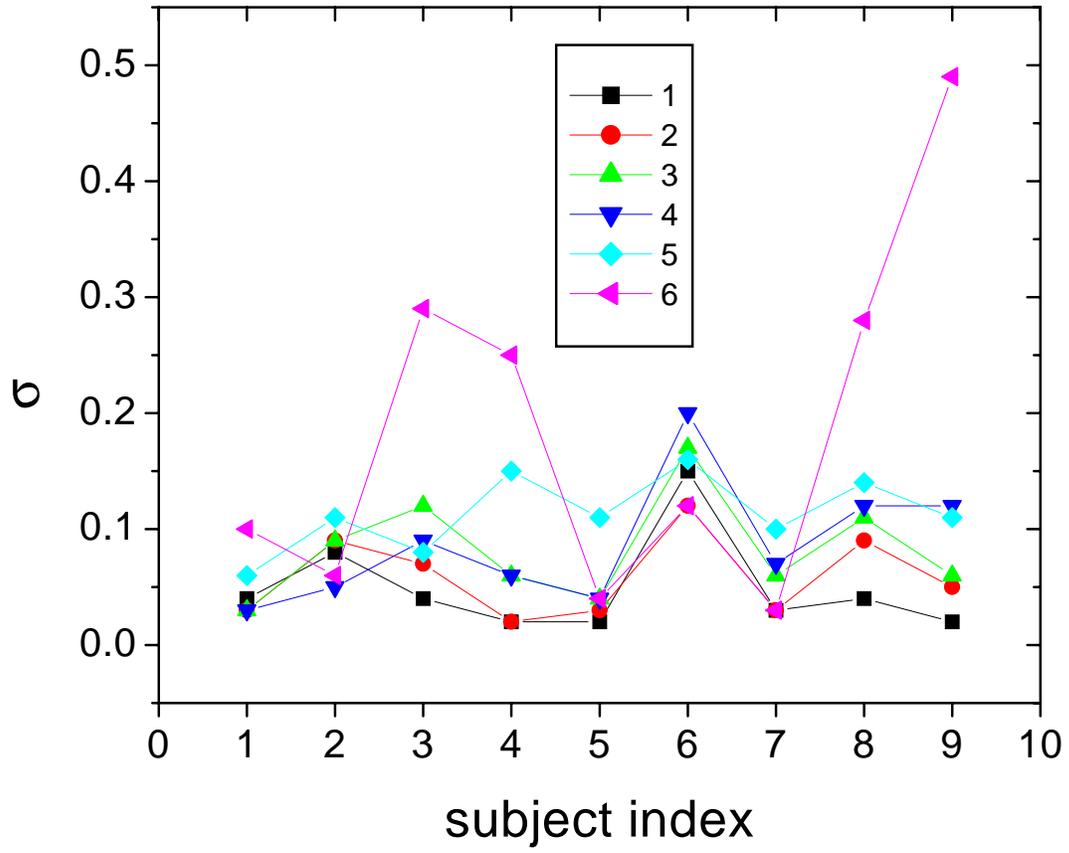,
bbllx=47 pt,bblly=334 pt,
bburx=537 pt,bbury=755 pt,
width=0.9\textwidth,clip=}}
\caption{
Standard deviation of the wavelet coefficients as a function of scale index
and subject index. The corresponding scale indexes are shown in the inset.
This result is  for the supine position.
}
\label{fig_4}
\end{figure}

\newpage
\begin{figure}[tbp]
\centerline{\epsfig{file=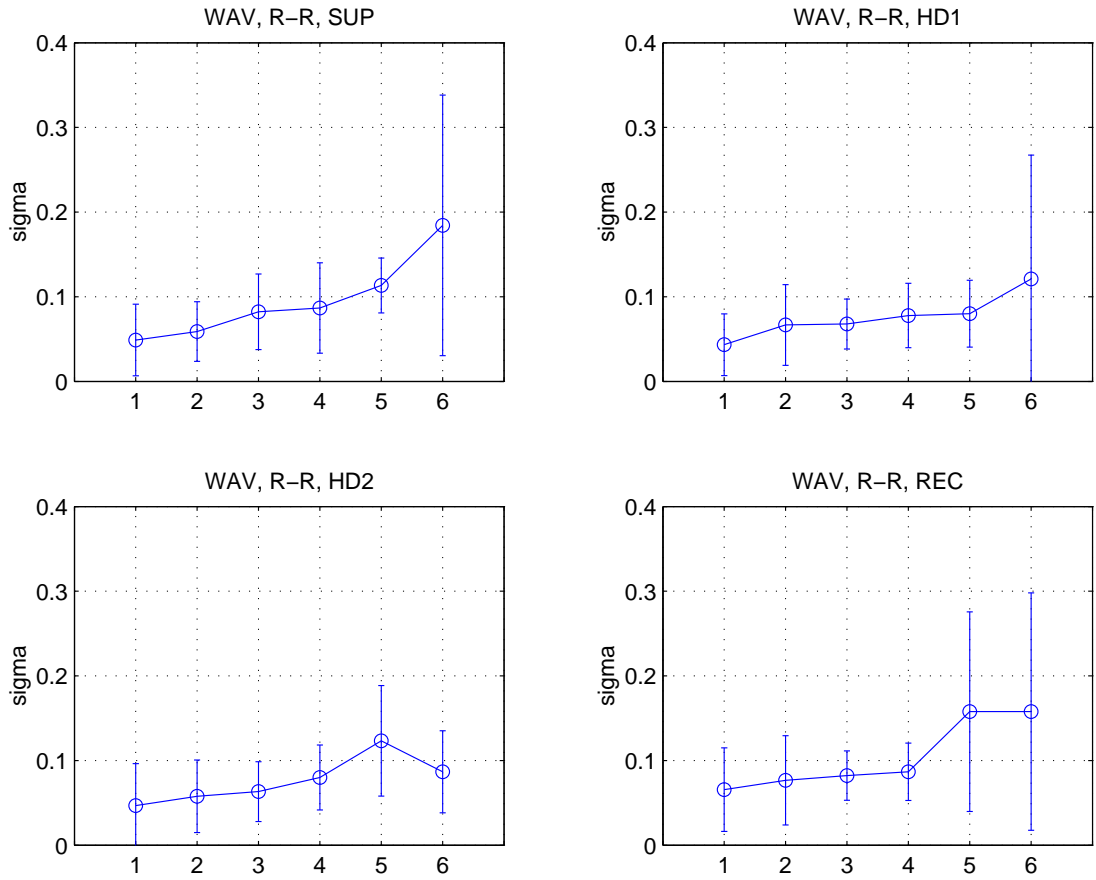,
bbllx=55 pt,bblly=205 pt,
bburx=545 pt,bbury=605 pt,
width=0.9\textwidth,clip=}}
\caption{
Mean value of the standard deviation of the wavelet coefficients with 
error-bars as a function of scale index $j$. This result is for the
R-R intervals.
}
\label{fig_5}
\end{figure}

\newpage
\begin{figure}[tbp]
\centerline{\epsfig{file=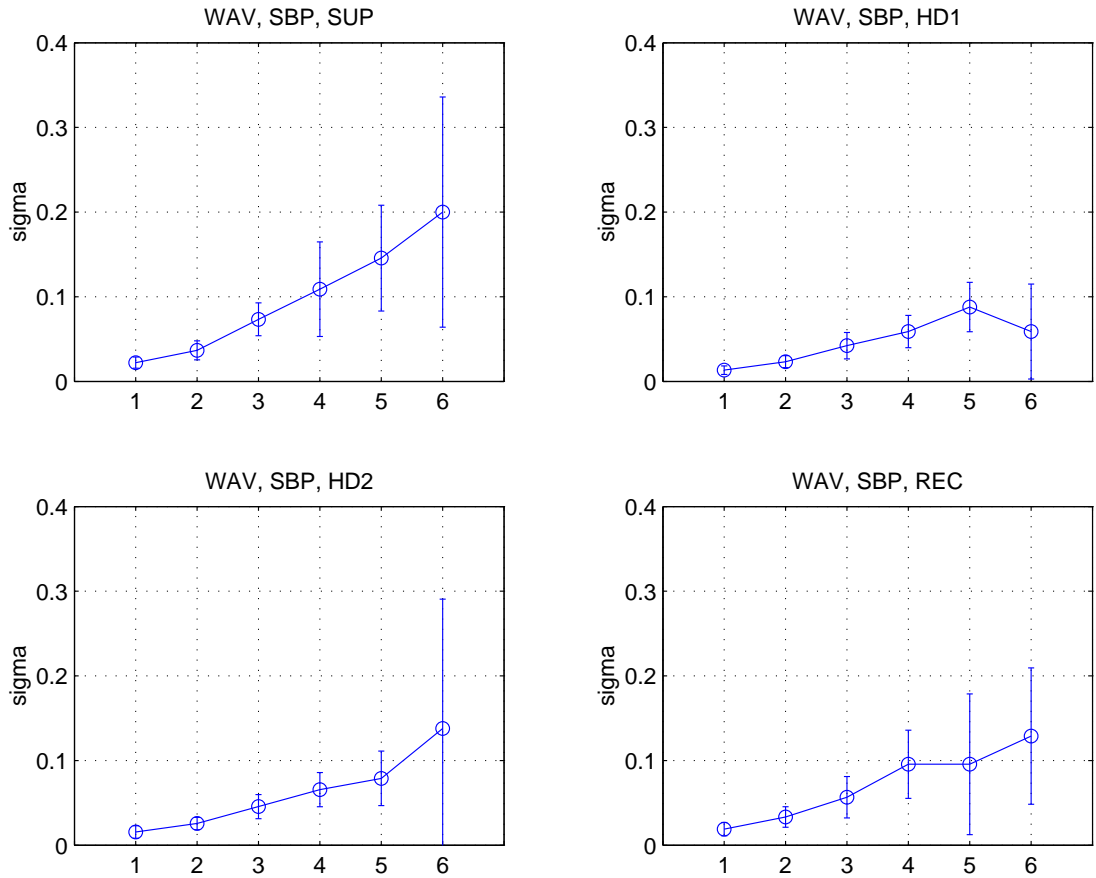,
bbllx=55 pt,bblly=205 pt,
bburx=545 pt,bbury=605 pt,
width=0.9\textwidth,clip=}}
\caption{
Mean value of the standard deviation of the wavelet coefficients with 
error-bars as a function of scale index $j$. This result is for the
systolic blood pressures.
}
\label{fig_6}
\end{figure}

\newpage
\begin{figure}[tbp]
\centerline{\epsfig{file=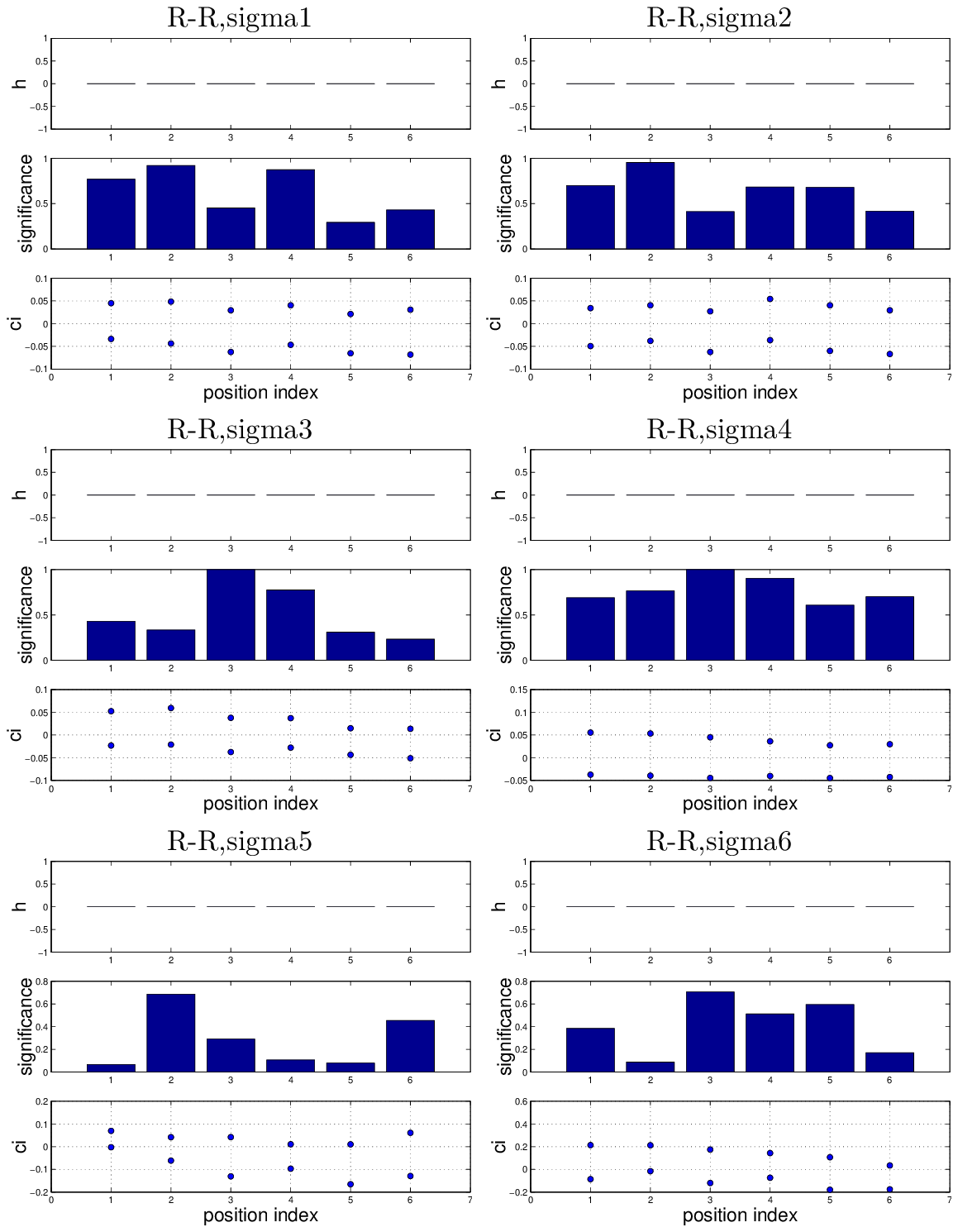,
bbllx=37 pt,bblly=211 pt,
bburx=475 pt,bbury=678 pt,
width=0.9\textwidth,clip=}}
\caption{
T-test-2 results of $\sigma$ of the six used scales 
for any position-pair of the 
four body positions. ``sigma1'', ..., ``sigma6'' in the figure
correspond to scale indexes 1, ..., 6, respectively.
 The position indexes
 1,2,3,4,5,6 correspond to the position-pairs SUP-HD1, SUP-HD2, SUP-REC, 
HD1-HD2, HD1-REC, HD2-REC, respectively.  
This result is for the R-R intervals.
}
\label{fig_7}
\end{figure}

\newpage
\begin{figure}[tbp]
\centerline{\epsfig{file=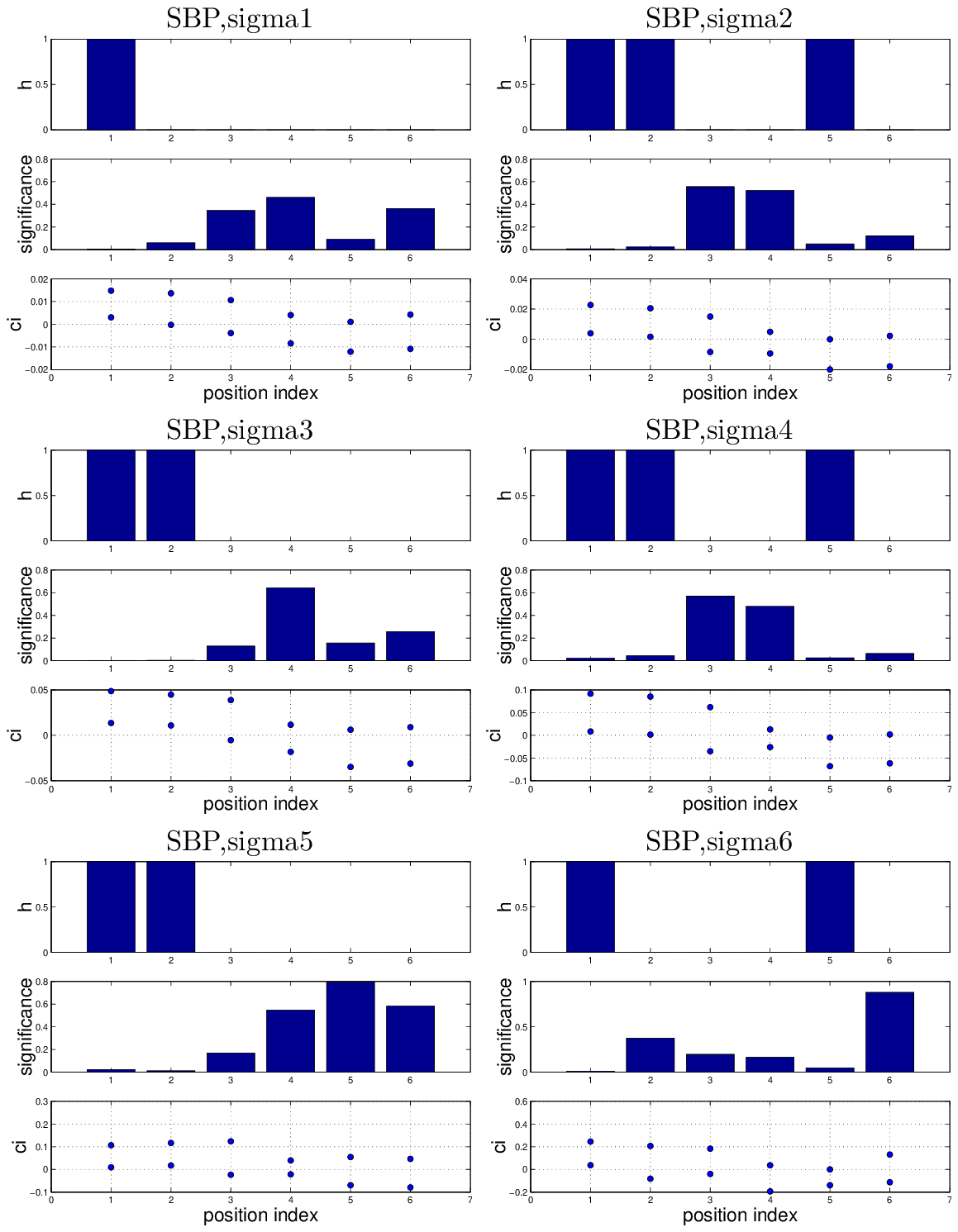,
bbllx=37 pt,bblly=211 pt,
bburx=475 pt,bbury=678 pt,
width=0.9\textwidth,clip=}}
\caption{
T-test-2 results of $\sigma$ of the six used scales 
for any position-pair of the 
four body positions. ``sigma1'', ..., ``sigma6'' in the figure
correspond to scale indexes 1, ..., 6, respectively.
 The position indexes
 1,2,3,4,5,6 correspond to the position-pairs SUP-HD1, SUP-HD2, SUP-REC, 
HD1-HD2, HD1-REC, HD2-REC, respectively.  
This result is for the systolic blood pressures.
}
\label{fig_8}
\end{figure}


\begin{references}
\bibitem{hamming}  A discrete description of the Hamming window function is $%
W(k+i)=0.54-0.46\cos [2\pi k/(n-1)]$, $k=0$, $\cdots $, $n-1$, where $n$ is
the width of window, $i$ is the index of the window.
\end{references}
\end{document}